\documentclass[12pt,dvips]{article}
\usepackage{amsfonts}
\usepackage{amsmath}
\usepackage{epsfig}
\usepackage{latexsym}
\usepackage{axodraw}
\newcommand{\nc}{\newcommand}
\nc{\postscript}[2]
{\setlength{\epsfxsize}{#2\hsize}\centerline{\epsfbox{#1}}}
\nc{\non}{\nonumber}
\nc{\hc}{\hbox {H.c.}} \nc{\re}{\hbox {Re}} 
\nc{\mev}{\hbox {MeV}} \nc{\gev}{\;\hbox {GeV}} \nc{\tev}{\;\hbox {TeV}}
\def\lsim{\mathrel{\raise.3ex\hbox{$<$\kern-.75em\lower1ex\hbox{$\sim$}}}}
\def\gsim{\mathrel{\raise.3ex\hbox{$>$\kern-.75em\lower1ex\hbox{$\sim$}}}}

\nc{\prd}[3]{{\it Phys.\ Rev.}\ {{\bf D{#1}} (#2), #3}}
\nc{\prl}[3]{{\it Phys.\ Rev.\ Lett.}\ {{\bf {#1}} (#2), #3}}
\nc{\plb}[3]{{\it Phys.\ Lett.}\ {{\bf B{#1}} (#2), #3}}
\nc{\npb}[3]{{\it Nucl.\ Phys.}\ {{\bf B{#1}} (#2), #3}}
\nc{\ptp}[3]{{\it Prog.\ Theor.\ Phys.}\ {{\bf {#1}} (#2), #3}}
\nc{\zfp}[3]{{\it Z.\ Phys.}\ {{\bf C{#1}} (#2), #3}}
\nc{\epj}[3]{{\it Eur.\ Phys.\ J.}\ {{\bf C{#1}} (#2), #3}}
\nc{\mpla}[3]{{\it Mod.\ Phys.\ Lett.}\ {{\bf A{#1}} (#2), #3}}
\nc{\rmp}[3]{{\it Rev.\ Mod.\ Phys.}\ {{\bf {#1}} (#2), #3}}
\nc{\ijmpa}[3]{{\it Int.\ J.\ of\ Mod.\ Phys.}\
               {{\bf A{#1}} (#2), #3}}
\nc{\Lsp}{\;\;\;\;\;\;\;\;\;\;}  \nc{\LLLsp}{\lspace \lspace}
\nc{\lsp}{\;\;\;\;\;\;}
\nc{\spac}{\;\;\;}
\nc{\noi}{\noindent}
\nc{\beq}{\begin{equation}}
\nc{\eeq}{\end{equation}}
\nc{\bea}{\begin{eqnarray}}
\nc{\eea}{\end{eqnarray}}
\nc{\baa}{\begin{array}}
\nc{\eaa}{\end{array}}
\nc{\bit}{\begin{itemize}}
\nc{\eit}{\end{itemize}}
\nc{\ben}{\begin{enumerate}}
\nc{\een}{\end{enumerate}}
\nc{\bce}{\begin{center}}
\nc{\ece}{\end{center}}

\def\sq2{\sqrt{2}}

\def\ph{\varphi}

\def\m4{m^4(\ph)}
\def\mn2{m_n^2}

\def\r0{|R_0|}

\newlength{\dinwidth}
\newlength{\dinmargin}
\DeclareMathAlphabet{\scr}{U}{rsfs}{m}{n}
\begin{document}
\pagestyle{empty}
\begin{flushright}
CERN-PH-TH/2005-027
\end{flushright}
\vskip 2cm
\begin{center}
{\Huge R-symmetries from higher dimensions}
\vspace*{5mm} \vspace*{1cm}
\end{center}
\vspace*{5mm} \noindent
\vskip 0.5cm
\centerline{\bf Zygmunt Lalak${}^{1,2}$
and Rados\l aw Matyszkiewicz${}^3$}
\vskip 1cm
\centerline{\em ${}^{1}$Institute of Theoretical Physics}
\centerline{\em University of Warsaw, Poland}
\vskip 0.3cm
\centerline{\em ${}^{2}$Theory Division, CERN}  
\centerline{\em CH-1211 Geneva 23, Switzerland}  
\vskip 0.3cm
\centerline{\em ${}^{3}$Institute of Nuclear Physics}
\centerline{\em ul. Radzikowskiego 152} 
\centerline{\em PL-31-342 Cracow, Poland}  
\vskip 1.5cm
\vskip1cm
\centerline{\bf Abstract}
\vskip 0.5cm
The supersymmetric extensions of the Standard Model can tolerate quite a large hierarchy 
between various supersymmetry breaking terms, a good example being the models of split supersymmetry.
However, theoretical models generating such a stable hierarchy are not so easy to construct. 
An interesting idea consists in coupling the brane localized gauge sector to extended supergravities in the bulk 
of extra dimensions, and using different sources of supersymmetry breaking in the bulk and on the brane. 
This in principle allows one to separate the magnitude of the gravitino mass from the supersymmetry 
breaking masses of gaugini and of charged matter. In this paper we present a detailed analysis of a simple 
field theoretical model where such an idea is realized. Departure from this symmetric set of boundary conditions
 breaks R-symmetry, and gaugino masses are generated at one-loop order, 
however the magnitude of the resulting soft gaugino masses is proportional to the R-symmetry breaking 
Majorana-type gravitino mass 
which is continously deformable to zero. 
\newpage
\pagestyle{plain}

\section{Introduction}
Supersymmetric field theories in four space-time dimensions are leading candidates to provide the next stage of 
unification of  fundamental 
interactions beyond the description offered by the Standard Model.  They are natural low-energy descendants 
of higher-dimensional fundamental theories such as superstrings and ten- or eleven-dimensional 
supergravities. However, supersymmetry must be broken at low energies, and understanding the pattern 
of masses and couplings which describes this breaking is  one of the central issues of the 
theory of fundamental interactions. Moreover, the  forthcoming experiments will be able to test TeV-scale supersymmetry breakdown,
hence exploration of theoretical possibilities leading to realistic predictions becomes more and more 
relevant.  One of the obvious questions which arises in this context is how large a hierarchy between 
the scales parametrizing supersymmetry breakdown is tolerable. 
This issue has been raised recently in a series of papers on split supersymmetry, \cite{Arkani-Hamed:2004fb,Giudice:2004tc}, where it has been found 
that among fermionic superpartners just light gauginos and higgsinos are sufficient to keep the model 
within experimental limits and to retain a number of interesting predictions. 

However, the question about a natural mechanism generating a hierarchy between supersymmetry breaking terms 
remains open. In general, within the framework of N=1 supergravity in 4d generating a significant hierarchy 
between supersymmetry breaking terms is problematic. This is more or less expected in the scenarios of gravity 
mediated supersymmetry breakdown, since the hidden sector breakdown is characterized by a single scale, 
like condensation scale of strong gauge dynamics, and mediation is modulated only by expectation values 
of moduli fields, which cannot differ to much as they are determined by the same potential which 
switches-on the supersymmetry breakdown. Another argument based on the particular structure 
of N=1 supergravity is the observation, that Majorana gaugini masses are forbidden by R-symmetry,
and this symmetry is broken when the gravitino becomes massive. This is because the N=1 gravitino mass 
term  arises by means of  the super-higgs effect. In fact, typically the fermions which supply the helicity 1/2 
components to the gravitino come from the chiral multiplets (we neglect general D-type
breaking as it needs a non-trivial F-component). As a consequence the gravitino mass term depends 
on the nonzero expectation value  of  the superpotential, which always breaks R-symmetry.

An interesting proposal which avoids this problem within the framework of higher-dimensional locally supersymmetric theories 
has been put forward in \cite{Antoniadis:2004dt}. There gravitini can obtain Dirac-type mass through mixing with additional 
degrees of freedom from the gravitational multiplet, which are there due to the N=2 superesymmetry in the bulk.
The interesting feature of this mechanism is that at tree level it decouples gravitino mass from the 
scale of the supersymmetry breaking in the gauge sector, which gives a hope for creating hierarchy between 
supersymmetry breaking masses. In addition, one can break the R-symmetry continously in the gravity sector,
using the brane terms, which is equivalent to adjusting continously boundary conditions. This breaking 
is communicated in loops to the gauge sector living on the branes. 

This mechanism has been analysed in detail at the level of string theory construction, \cite{Antoniadis:2004dt}. 
In this paper we give a detailed description at the level of five-dimensional supergravity. 
It is interesting to note, that the case with broken supersymmetry but exact R-symmetry corresponds 
precisely to flipped supergravity of \cite{Brax:2001xf,Lalak:2002kx,Lalak:2003fu}.   

To illustrate the difference between four-dimensional and higher-dimensional superhiggs effects, at the end of the paper 
we describe the superhiggs effect  in N=2 supergravity with flipped boundary conditions.

\section{R--symmetry breakdown}
The model we discuss here is the simple $N=2$ five-dimensional supergravity with branes, basic features of which 
are summarized in the appendix. 
In this paper we consider for simplicity flat geometry, 
hence we do not put in any cosmological term in the bulk, 
and there are no explicit brane tensions at the fixed points. 
However, one can freely enhance the model by non-zero gravitini masses localized on the branes, which fully respect 
five--dimensional supersymmetry (for details see \cite{Lalak:2003fu}). Thus, the  relevant boundary Lagrangian can be written
as
	\begin{equation} \label{branelagranggeneral}
        e_4 {\cal L}_{brane}=
        -\sum_i e_4\delta(y-y_i)\bar{\Psi}_\mu^A\gamma^{\mu\nu}(M_i+\gamma_5\bar{M}_i)_{A}^{\;B} \Psi_{\nu B} \ ,
    \end{equation}
     where $(M_i)_{AB}$ and $(\bar{M}_i)_{AB}$ are symmetric matrices which denote
    gravitini masses on the branes. In addition, one needs to  modify the supersymmetry transformation of the fifth 
component of the gravitino
	\begin{equation}
        \delta\Psi^A_5 \longrightarrow\delta\Psi^A_5
        +2\delta_\alpha^{\;5}\sum_i \delta(y-y_i)(M_i+\gamma_5\bar{M}_i)^{A}_{\;B}\gamma_5\eta^B\ .
    \end{equation}

	The N=2 five-dimensional supergravity is invariant under the $SU(2)_R$ symmetry. Metric tensor and graviphoton form
	singlets, while gravitini and parameters of supersymmetry transformations form doublets with respect to this symmetry.
	One can check that the orbifold projections at the given brane, generated by the ${\bf Z_2}$ operator  
	and the gravitini masses on the brane, 
	break $SU(2)_R$ symmetry to a $U(1)$ subgroup. If the projections  breaks R--symmetry 
	to the same subgroup at each brane, the $U(1)_R$ symmetry remains unbroken in the effective theory. In the other case 
            all generators
	of the $SU(2)_R$ symmetry are broken. To be more specific, let us find explicitly the unbroken generator for a given 
            projection.
	
	Let us assume  equal ${\bf Z_2}$ operators on both branes: $Q_0=Q_\pi=\sigma_3$. In addition, let us allow  
	(locally on each brane) only the even  
		components of gravitini to have localized mass terms, i.e. 
    \begin{eqnarray}\label{gravitinimasses2}
        &&(M_{0})_A^{\;B}=\frac{1}{2}\alpha_{0}(\sigma_1)_A^{\;B}\ ,\qquad 
		(M_{\pi})_A^{\;B}=\frac{1}{2}\alpha_{\pi}(\sigma_1)_A^{\;B}\ ,\nonumber\\
        &&(\bar{M}_{0})_A^{\;B}=\frac{1}{2}{\rm i}\alpha_{0}(\sigma_2)_A^{\;B}\ ,\qquad
        (\bar{M}_{\pi})_A^{\;B}=\frac{1}{2}{\rm i}\alpha_{\pi}(\sigma_2)_A^{\;B}\ ,
    \end{eqnarray}
    where $\alpha_{0,\pi}$ are real coefficients\footnote{
In general, such terms can also be generated by the condensition of the superpotentials localized on the branes,  $\alpha_i=\langle W_i\rangle$.}.
    
	Then the boundary conditions for the gravitini read
      \begin{eqnarray} \label{warunkibrzeggrav2}
      &&\epsilon^{-1}(y) \delta(y) \gamma_5(\Psi_-)_\mu^A
  =-\delta(y)\alpha_{0}\sigma_1(\Psi_+)_\mu^A\
  ,\nonumber\\
    &&\epsilon^{-1}(y) \delta(y-\pi r_c) \gamma_5(\Psi_-)_\mu^A
  =\delta(y-\pi r_c)\alpha_{\pi}\sigma_1(\Psi_+)_\mu^A\ ,
      \end{eqnarray}      
      where we have decomposed the gravitini into
		the ${\bf Z_2}$-even $(+)$ and ${\bf Z_2}$-odd $(-)$ components 
		\begin{equation} \label{defpmspinors}
        (\Psi_{\pm})^A_\alpha=\frac{1}{2}(\delta\pm\gamma_5\sigma_{3})^{A}_{\;B}\Psi_\alpha^B\ .
		\end{equation}
        
		One can check that the unique $U(1)$ subgroup that leaves the boundary conditions invariant is generated
		by
		\begin{equation}\label{unbgen0}
		P^A_{\;B}=\left(\frac{2\alpha_0}{\alpha_0^2+1}\;\sigma_1
	+\frac{\alpha_0^2-1}{\alpha_0^2+1}\;\sigma_3\right)^A_{\;B}\ ,
		\end{equation}
	for the projection acting on the $y=0$ brane, and by 
	\begin{equation} \label{unbgenpi}
		P^A_{\;B}=\left(\frac{-2\alpha_\pi}{\alpha_\pi^2+1}\;\sigma_1
	+\frac{\alpha_\pi^2-1}{\alpha_\pi^2+1}\;\sigma_3\right)^A_{\;B}\ ,
		\end{equation}
	for the brane at $y=\pi r_c$. It is important to note  that the brane action localized at each brane {\it is  not} invariant 
on its own under the respective unbroken $U(1)_R$ symmetry. 
To see the invariance of the full brane plus bulk action, one needs to include the relevant contributions 
from the five-dimensional bulk action (see \cite{Lalak:2003cs}).

	\section{General solution of the gravitini equation of motion and compactification}	 
To compactify the model to four dimensions one needs to solve 5d equations of motion for gravitini.  
They  take the following form in the bulk 
    \begin{eqnarray} \label{equationosmo2}
    &&\gamma^{\mu\nu\rho}\partial_\nu\Psi_{\rho}^{A}-
	\gamma^5\gamma^{\mu\nu}\partial_5\Psi_{\nu}^{A}=0\ ,\nonumber\\ 
	&&\gamma^5\gamma^{\mu\nu}\partial_\mu\Psi_{\nu}^{A}=0\ ,
    \end{eqnarray}
    where we have chosen the gauge $\Psi_5=0$. The boundary conditions are given by (\ref{warunkibrzeggrav2}).
	The solution can be expressed as a linear combination of 
    the sine and cosine functions
        \begin{eqnarray} \label{solgravitini}
        &(\Psi_{+})_{\mu}^A&=\sum_n A^{(n)}\cos(m_n|y|)\left(\begin{array}{c}\psi^{(n)}_{\mu\,R}\\
        \chi^{(n)}_{\mu\,L}\end{array}\right)^A+\sum_n B^{(n)}\sin(m_n|y|)
		\left(\begin{array}{c}\hat{\psi}^{(n)}_{\mu\,R}\\
        \hat{\chi}^{(n)}_{\mu\,L}\end{array}\right)^A\nonumber\\
        &(\Psi_{-})_{\mu}^A&=\epsilon(y) \sum_n A^{(n)} \sin(m_n|y|)\left(\begin{array}{c}-\psi^{(n)}_{\mu\,L}\\
        \chi^{(n)}_{\mu\,R}\end{array}\right)^A+\epsilon(y) \sum_n B^{(n)}\cos(m_n|y|)\left(\begin{array}{c}\hat{\psi}^{(n)}_{\mu\,L}\\
        -\hat{\chi}^{(n)}_{\mu\,R}\end{array}\right)^A
        \end{eqnarray}
        where $\psi^{(n)}_\mu$, $\hat{\psi}^{(n)}_\mu$, $\chi^{(n)}_\mu$, $\hat{\chi}^{(n)}_\mu$ denote
        4d gravitini  in the flat space, which satisfy
        \begin{eqnarray} \label{equationofmgn5}
        &&\gamma^{\mu\rho\nu}\partial_{\rho}\psi^{(n)}_{\nu}
		- m_n\,\gamma^{\mu\nu}\psi^{(n)}_{\nu}=0\nonumber\\
		&&\gamma^{\mu\rho\nu}\partial_{\rho}\chi^{(n)}_{\nu}
		- m_n\,\gamma^{\mu\nu}\chi^{(n)}_{\nu}=0\ ,
		\end{eqnarray}
		 with additional conditions $\gamma^{\mu\nu}\partial_\mu\psi^{(n)}_{\nu}=0$ and 
		$\gamma^{\mu\nu}\partial_\mu\chi^{(n)}_{\nu}=0$. The symplectic Majorana condition implies 
		$\bar{\psi}^{(n)}_\mu=(\chi^{(n)}_\mu)^T C$ (hatted spinors share the same properties).  
		
		The boundary condition (\ref{warunkibrzeggrav2}) at the point $y=0$
		 implies $B^{(n)}\hat{\psi}_\mu^{(n)}=\alpha_0 A^{(n)}\chi_\mu^{(n)}$ and 
		$B^{(n)}\hat{\chi}_\mu^{(n)}=\alpha_0 A^{(n)}\psi_\mu^{(n)}$, hence one can write 
		\begin{eqnarray} \label{solgravitini2}
        &(\Psi_{+})_{\mu}^A&=\sum_n A^{(n)}\left(\cos(m_n|y|)\left(\begin{array}{c}\psi^{(n)}_{\mu\,R}\\
        \chi^{(n)}_{\mu\,L}\end{array}\right)^A+\alpha_0\sin(m_n|y|)
		\left(\begin{array}{c}\chi^{(n)}_{\mu\,R}\\
        \psi^{(n)}_{\mu\,L}\end{array}\right)^A\right)\nonumber\\
        &(\Psi_{-})_{\mu}^A&=\epsilon(y) \sum_n A^{(n)} \left(\sin(m_n|y|)\left(\begin{array}{c}-\psi^{(n)}_{\mu\,L}\\
        \chi^{(n)}_{\mu\,R}\end{array}\right)^A+ \alpha_0 \cos(m_n|y|)
		\left(\begin{array}{c}\chi^{(n)}_{\mu\,L}\\
        -\psi^{(n)}_{\mu\,R}\end{array}\right)^A\right)\ .
        \end{eqnarray}
        
		The boundary condition at $y=\pi r_c$ implies in turn 
		\begin{eqnarray} \label{bound2}
        &&(1-\alpha_0\alpha_\pi)\sin(m_n\pi r_c)\psi_{\mu\,L}^{(n)}=(\alpha_0+\alpha_\pi)\cos(m_n\pi r_c)\chi_{\mu\,L}^{(n)}\nonumber\\
        &&(1-\alpha_0\alpha_\pi)\sin(m_n\pi r_c)\chi_{\mu\,R}^{(n)}=(\alpha_0+\alpha_\pi)\cos(m_n\pi r_c)\psi_{\mu\,R}^{(n)}\ .
        \end{eqnarray}
		
		We shall  solve these equations considering separately various  cases for the gravitini masses.
		\begin{itemize}
		\item Let us start with $\alpha_0=\alpha_\pi=0$\\\vspace{0.3cm}\\
		 The condition (\ref{bound2}) gives the following  quantization of the masses:
		\begin{equation}
		\sin(m_n\pi r_c)=0 \Longrightarrow m_n=\frac{{\rm n}}{r_c}\ ,
		\end{equation}
		where ${\rm n}=0,1,2,...$
		The zero mode does exist and supersymmetry remains unbroken. The solution (\ref{solgravitini2}) takes
	 the form 
		 \begin{eqnarray} \label{solgravitinisp1}
        &(\Psi_{+})_{\mu}^A&=\sum_n A^{(n)}\cos(m_n|y|)\left(\begin{array}{c}\psi^{(n)}_{\mu\,R}\\
        \chi^{(n)}_{\mu\,L}\end{array}\right)^A\nonumber\\
        &(\Psi_{-})_{\mu}^A&=\epsilon(y) \sum_n A^{(n)}\sin(m_n|y|)\left(\begin{array}{c}-\psi^{(n)}_{\mu\,L}\\
        \chi^{(n)}_{\mu\,R}\end{array}\right)^A
        \end{eqnarray}
        and it is invariant under the symmetry $U(1)\subset SU(2)_R$ generated by  $(-\sigma_3)^A_{\;B}$.   
        The gravitini $\psi^{(n)}_\mu$ have a negative charge, say $-1$, with respect to this symmetry while 
		the ones denoted by $\chi^{(n)}_\mu$ have a positive charge $+1$. In fact, ona can check that the original $SU(2)_R$ symmetry is 
        broken down
        to this $U(1)$ subgroup by the boundary conditions imposed on the brane. We have obtained the {\em Dirac} masses in the effective theory,
		hence the effective four-dimensional action is invariant under the $U(1)_R$ symmetry related to  the unbroken $N=1$
		supersymmetry.
		\item In the second step let us discuss the case $\alpha_0=-\alpha_\pi$. \\\vspace{0.3cm}\\
		Again, the boundary conditions imply
		\begin{equation}
		\sin(m_n\pi r_c)=0 \Longrightarrow m_n=\frac{{\rm n}}{r_c}\ ,
		\end{equation}
		 and the solution
		 \begin{eqnarray} \label{solgravitinisp2}
        &(\Psi_{+})_{\mu}^A&=\sum_n A^{(n)}\Bigg(\cos(m_n|y|)\left(\begin{array}{c}\psi^{(n)}_{\mu\,R}\\
        \chi^{(n)}_{\mu\,L}\end{array}\right)^A+\alpha_0\sin(m_n|y|)
		\left(\begin{array}{c}\chi^{(n)}_{\mu\,R}\\
        \psi^{(n)}_{\mu\,L}\end{array}\right)^A\Bigg)\nonumber\\
        &(\Psi_{-})_{\mu}^A&=\epsilon(y)\sum_n  A^{(n)} \Bigg(\sin(m_n|y|)\left(\begin{array}{c}-\psi^{(n)}_{\mu\,L}\\
        \chi^{(n)}_{\mu\,R}\end{array}\right)^A\nonumber\\&&\qquad\qquad\qquad
\qquad\qquad\qquad\qquad
		+ \alpha_0 \cos(m_n|y|)
		\left(\begin{array}{c}\chi^{(n)}_{\mu\,L}\\
        -\psi^{(n)}_{\mu\,R}\end{array}\right)^A\Bigg)
        \end{eqnarray}
		 preserves $N=1$ supersymmetry. Again,  the $U(1)\subset SU(2)_R$   survives compactification.  
		The unbroken generator is given by (\ref{unbgen0}). 
		Gravitini $\psi^{(n)}_\mu$ are negatively  charged, while 
		 $\chi^{(n)}_\mu$ have are positively charged with respect to this symmetry,  and we have 
obtained the {\em Dirac} mass terms in the effective theory.

		\item The choice $\alpha_0=1/\alpha_\pi$  corresponds to the flipped supergravity.\\\vspace{0.3cm}\\
		The boundary conditions imply
		\begin{equation}
		\cos(m_n\pi r_c)=0 \Longrightarrow m_n=\frac{{\rm n+\frac{1}{2}}}{r_c}\ ,
		\end{equation}
		 and the solution of the equations of motion is
		 \begin{eqnarray} \label{solgravitinisp3}
        &(\Psi_{+})_{\mu}^A&=\sum_n A^{(n)}\Bigg(\cos(m_n|y|)\left(\begin{array}{c}\psi^{(n)}_{\mu\,R}\\
        \chi^{(n)}_{\mu\,L}\end{array}\right)^A+\alpha_0\sin(m_n|y|)
		\left(\begin{array}{c}\chi^{(n)}_{\mu\,R}\\
        \psi^{(n)}_{\mu\,L}\end{array}\right)^A\Bigg)\nonumber\\
        &(\Psi_{-})_{\mu}^A&=\epsilon(y) \sum_n A^{(n)} \Bigg(\sin(m_n|y|)\left(\begin{array}{c}-\psi^{(n)}_{\mu\,L}\\
        \chi^{(n)}_{\mu\,R}\end{array}\right)^A
         \nonumber\\
&&\qquad\qquad\qquad
\qquad\qquad\qquad\qquad
		+ \alpha_0 \cos(m_n|y|)
		\left(\begin{array}{c}\chi^{(n)}_{\mu\,L}\\
        -\psi^{(n)}_{\mu\,R}\end{array}\right)^A\Bigg).
        \end{eqnarray}
        In this case supersymmetry is broken by the boundary conditions, nevertheless  the $U(1)\subset SU(2)_R$
		 symmetry
	    remains unbroken and
		the unbroken generator is given by (\ref{unbgen0}). 
		Again, the gravitini $\psi^{(n)}_\mu$ have the negative charge, while 
		the $\chi^{(n)}_\mu$ have the positive charge, and we have obtained the {\em Dirac} mass terms  in the effective 
theory.

		 
		\item Finally, we shall treat the remaining cases.\\\vspace{0.3cm}\\
		To solve the boundary conditions (\ref{bound2}), one needs to change the basis of the four-dimensional gravitini to:
		 \begin{equation}
		 \tilde{\psi}_\mu^{(n)}=\frac{1}{\sqrt{2}}\left(\psi^{(n)}_\mu+\chi^{(n)}_\mu\right)\ ,\qquad 
		\tilde{\chi}^{(n)}_\mu=\frac{1}{\sqrt{2}}\left(\psi^{(n)}_\mu-\chi^{(n)}_\mu\right)\ .
		 \end{equation}
		Then, the  equation (\ref{bound2}) reads
		\begin{eqnarray} \label{bound3}
		&&(1-\alpha_0\alpha_\pi)\sin(m_n\pi r_c)\tilde{\psi}^{(n)}_\mu=(\alpha_0+\alpha_\pi)\cos(m_n\pi r_c)\tilde{\psi}^{(n)}_\mu\nonumber\\
        &&(1-\alpha_0\alpha_\pi)\sin(m_n\pi r_c)\tilde{\chi}^{(n)}_\mu=-(\alpha_0+\alpha_\pi)\cos(m_n\pi r_c)\tilde{\chi}^{(n)}_\mu\ ,
        \end{eqnarray}
		which eventually leads to the following 
		quantization of the KK masses:
		\begin{eqnarray}
		&&m_{\tilde{\psi}}=\frac{1}{r_c}\left({\rm n}+\frac{1}{\pi}\arctan\left(\frac{\alpha_0+\alpha_\pi}{1-\alpha_0\alpha_\pi}
		\right)\right) \ ,\quad \;\;\;\;\;\;{\rm for}\;\;\;\frac{\alpha_0+\alpha_\pi}{1-\alpha_0\alpha_\pi}\geq 0\ ,\nonumber\\
		&&m_{\tilde{\chi}}=\frac{1}{r_c}\left({\rm n}+1-\frac{1}{\pi}\arctan\left(\frac{\alpha_0+\alpha_\pi}{1-\alpha_0\alpha_\pi}
		\right)\right) \ ,\quad {\rm for}\;\;\;\frac{\alpha_0+\alpha_\pi}{1-\alpha_0\alpha_\pi}\geq 0\ ,\nonumber\\
		&&m_{\tilde{\psi}}=\frac{1}{r_c}\left({\rm n}+1+\frac{1}{\pi}\arctan\left(\frac{\alpha_0+\alpha_\pi}{1-\alpha_0\alpha_\pi}
		\right)\right) \ ,\quad {\rm for}\;\;\;\frac{\alpha_0+\alpha_\pi}{1-\alpha_0\alpha_\pi}<0\ ,\nonumber\\
		&&m_{\tilde{\chi}}=\frac{1}{r_c}\left({\rm n}-\frac{1}{\pi}\arctan\left(\frac{\alpha_0+\alpha_\pi}{1-\alpha_0\alpha_\pi}
		\right)\right) \ ,\quad \;\;\;\;\;\;{\rm for}\;\;\;\frac{\alpha_0+\alpha_\pi}{1-\alpha_0\alpha_\pi}<0\ .
		\end{eqnarray}
	
		The solution takes the form
		\begin{eqnarray} \label{solgravitini3}
        &(\Psi_{+})_{\mu}^A&=\sum_n A^{(n)}_{\tilde{\psi}}\left(\cos(m_{\tilde{\psi}}|y|)+\alpha_0\sin(m_{\tilde{\psi}}|y|)\right)
		\left(\begin{array}{c}\tilde{\psi}^{(n)}_{\mu\,R}\\
        \tilde{\psi}^{(n)}_{\mu\,L}\end{array}\right)^A\nonumber\\
		&&+\sum_n A^{(n)}_{\tilde{\chi}}\left(\cos(m_{\tilde{\chi}}|y|)-\alpha_0\sin(m_{\tilde{\chi}}|y|)\right)\left(\begin{array}{c}\tilde{\chi}^{(n)}_{\mu\,R}\\
        -\tilde{\chi}^{(n)}_{\mu\,L}\end{array}\right)^A
		\ ,\nonumber\\
        &(\Psi_{-})_{\mu}^A&=\epsilon(y)\sum_n A^{(n)}_{\tilde{\psi}}\left(\alpha_0\cos(m_{\tilde{\psi}}|y|)-\sin(m_{\tilde{\psi}}|y|)\right)
		\left(\begin{array}{c}\tilde{\psi}^{(n)}_{\mu\,L}\\
        -\tilde{\psi}^{(n)}_{\mu\,R}\end{array}\right)^A\nonumber\\
		&&-\epsilon(y)\sum_n A^{(n)}_{\tilde{\chi}}\left(\alpha_0\cos(m_{\tilde{\chi}}|y|)+\sin(m_{\tilde{\chi}}|y|)\right)
		\left(\begin{array}{c}\tilde{\chi}^{(n)}_{\mu\,L}\\
        \tilde{\chi}^{(n)}_{\mu\,R}\end{array}\right)^A
		\ .
        \end{eqnarray}
       In this case supersymmetry is broken and the orbifold projections break down $SU(2)_R$ symmetry
		to different subgroups at different  branes, hence no $U(1)$ invariance survives in the effective theory. 
In particular, non-vanishing Majorana mass terms for gravitini are generated. 
 
        \end{itemize}

\section{Limitations of  four-dimensional description}
	            Let us  recall that in the effective theory, at energies
		below the compactification scale, one observes the zero modes
		of the particles that form N=1 massless supergravity multiplet and
		N=1 chiral supermultiplet.
		The effective N=1 supersymmetric action is determined by a
		K\"ahler potential $K$ and a superpotential $W$.  
		Reduction of the five-dimensional bosonic action in the flat case leads to the
		following form of the K\"ahler function
		\begin{equation}
		K=-3\log\left(T+\bar{T}\right)\ ,
		\end{equation}
		where
		\begin{equation}
		T= r+{\rm i}A\ ,
		\end{equation}
                        $r$ denotes the proper radius of the fifth dimension in the original 5d coordinates and
		$A$ denotes the axion.
		The only form of the superpotential that leads to the vanishing scalar potential is a constant
		$W=\omega $.
                        In the flat compactification performed in the previous section 
		the effective scalar potential vanishes. As a consequence, the proper radius of the fifth
		dimension, hence, the vacuum expectation value of the $T$ field is undetermined. 
		In the previous section we have denoted  the proper radius by $r_c$, assuming that there exists some mechanism
		(in fact unknown) which determines  this  value. Then we performed rescaling of the fifth coordinate that the 
		expectation value of the $e^{\hat{5}}_5$ is 1. In such a case the curvature scalar in five and four dimensions are 
		equal and one do not need the Weyl rescaling, turning from five-dimensional to four-dimensional theory. 
		
		Now, we would like to keep the freedom of choosing the vacuum expectation value of the $T$ field. Hence, we assume that
		the value of $\langle e^{\hat{5}}_5\rangle$ is undetermined. Then the proper radius is $r_c \langle e^{\hat{5}}_5\rangle$.
		It is more convenient to put the value of $r_c$ equal to 1 (now, it is only a free parameter) and identify
		the proper radius ($T$ field) with  $\langle e^{\hat{5}}_5\rangle$. To obtain the canonical curvature scalar in
		four dimensions, the following Weyl rescaling is needed:
		 \begin{equation}
		 g_{\mu\nu}\longrightarrow \langle r\rangle^{-1} g_{\mu\nu}\ .
		 \end{equation}
		 Then the  mass of the lowest Kaluza-Klein mode of the gravitino changes to\footnote{In fact, one can argue 
                         that  $\langle r\rangle^{-\frac{3}{2}}$ gives the effective physical radius of the fifth dimension,
                        see \cite{Bucci:2004xw}.}  
		 \begin{equation}\label{gravmas}
		m_{3/2}=\frac{1}{2r_c}\langle r\rangle^{-\frac{3}{2}}=\frac{1}{2}\langle r\rangle^{-\frac{3}{2}}\ .
		\end{equation} 
                        In this paper we mostly use, for convenience, mass terms corresponding to  the 5d canonical normalization
                        of the gravitational action, however, the need for the final Weyl rescaling is always understood. 
 
		In the
		four-dimensional supergravity gravitino mass is proportional to the
		vacuum expectation value of the 4d superpotential
		\begin{equation}
		m_{3/2}=\langle e^{\frac{K}{2}} W\rangle\ ,
		\end{equation}
		The calculations made in  our effective four-dimensional
		model lead to
		\begin{equation}
		m_{3/2}=\omega\langle T+\bar{T}\rangle^{-\frac{3}{2}}=\omega \langle 2 r\rangle^{-\frac{3}{2}}
		\end{equation}
		and agree with the five-dimensional gravitini mass of the lowest Kaluza-Klein mode (\ref{gravmas})
		for $\omega=\sqrt{2}$.
		
		One can calculate the vacuum expectation value of the superpotential which leads to spontaneously 
		broken supersymmetry in the effective four-dimensional supergravity, for a given set
		of boundary conditions in five-dimensional models:  
		\begin{equation}
		W=\frac{2\sqrt{2}}{\pi}\arctan\left(\frac{\alpha_0+\alpha_\pi}{1-\alpha_0\alpha_\pi}
		\right)\ .
		\end{equation}
		Note that the four-dimensional supergravity presented above describes effective theory at energies
		below the compactification scale, where one observes the lightest modes
		of the particles. In fact, this  formalism can be valid for the scale of supersymmetry breaking
		much smaller than the comactification scale. In the other case a gap between the  masses of the first and second
		 Kaluza-Klein states of the gravitini  is relatively small and it is difficult to find a proper scale below which
		 one observes only the lightest gravitino. In the limiting case (flipped supergravity) these masses are equal, hence
	   above four-dimensional description totally breaks down. We have obtained a novel and unique four-dimensional theory
	   that consists of one massless graviton and two massive gravitini. In addition, the mass terms for the gravitini 
		in the effective 
		Lagrangian are of the {\em Dirac} type, hence they are invariant under the $U(1)_R$ symmetry.

		\section{Coupling to the matter localized on the branes}
		In the N=1 four-dimensional supergravity  left and right components of the
	    gravitino have opposite charges with respect to the $U(1)_R$ symmetry. The complete theory including gauge fields and chiral matter 
can be arranged to be invariant under the $R$-symmetry by the apropriate choice of the superpotential, and the gravitational sector is invariant
		under this symmetry because  gravitino mass terms, which in principle could break it, are absent.
		 However, gravitino couples to the  superpotential:
		\begin{equation}
		{\cal L}_4\supset W(\Phi,\bar{\Phi})\bar{\psi_\mu}\gamma^{\mu\nu}\psi_\nu\ ,
		 \end{equation}
		and the nonzero vacuum expectation value of the superpotential spontaneously breaks supersymmetry as well as 
		R--symmetry. Effectively, one obtains the {\em Majorana} masses for the chiral gravitini $m_{eff}\sim\langle W\rangle$.   	
		In the matter and gauge sectors, supersymmetry breaking manifests itself through masses of  scalars and masses of gaugini.
		 The first ones arise at  tree
	   level from the explicit coupling to the F--terms ${\cal L}_4\supset |F|^2 \Phi^2$, where
		\begin{equation}
		F^{i}=K^{i\bar{j}}D_{\bar{j}}\bar{W}e^{K/2}\ ,
		\end{equation}
		and are of the same order as the gravitino mass.
		The masses of gaugini are generated by  loop corrections. To be more specific let us consider the relevant coupling 	
		in the four--dimensional supergravity 
		 \begin{equation}
		 {\cal L}_4\supset -\frac{1}{4}\bar{\psi}_\mu\gamma^{\nu\rho}\gamma^\mu\lambda\,\bar{\psi}_\nu\gamma^{\rho}\lambda\ .
		 \end{equation}
		 The Fierz rearrangement leads to the following form useful for loop calculations
		 \begin{eqnarray}\label{couplgg}
		 &&{\cal L}_4\supset \frac{1}{16}\bar{\psi}_\mu\gamma^{\mu\nu}(1+\gamma_5)\psi_\nu\,\bar{\lambda}(1-\gamma_5)\lambda+\frac{1}{16}\bar{\psi}_\mu\gamma^{\mu\nu}(1-\gamma_5)\psi_\nu\,\bar{\lambda}
		(1+\gamma_5)\lambda\nonumber\\
		 &&\;\;\,\quad-\frac{3}{16}\bar{\psi}_\mu(1+\gamma_5)\psi^\mu\,\bar{\lambda}(1-\gamma_5)\lambda-\frac{3}{16}\bar{\psi}_\mu(1-\gamma_5)\psi^\mu\,\bar{\lambda}
		(1+\gamma_5)\lambda\nonumber\\
		&&\;\;\,\quad+\frac{1}{8}\bar{\psi}_\mu\gamma^{\mu}\gamma_5\psi_\nu\,\bar{\lambda}\gamma^\nu\gamma_5\lambda+\frac{1}{16}\bar{\psi}_\mu\gamma^\rho\gamma_5\psi^\mu\,\bar{\lambda}\gamma_\rho\gamma_5\lambda-\frac{\rm i}{16}\bar{\psi}_\mu\epsilon^{\mu\rho\nu\sigma}\gamma_\sigma\psi_\nu\,\bar{\lambda}
		\gamma_\rho\gamma_5\lambda\ .
		 \end{eqnarray}
		 
       One can check that only terms in the two first lines in (\ref{couplgg}) can  contribute to the effective mass terms for gaugini.

		\begin{figure}[htbp]
\begin{center}
\begin{picture}(400,80)(0,0)

\ArrowLine(10,10)(89,10)
\ArrowLine(91,10)(170,10)
\PhotonArc(90,40)(30,0,360){2}{30}
\ArrowArc(90,40)(30,-90,90)
\ArrowArc(90,40)(30,90,270)
\Vertex(90,10){1.5} \Vertex(90,70){2}
\Text(140,40)[]{$\psi^{\mu}_{R}$}
\Text(40,40)[]{$\bar{\psi}^{\nu}_{L}$}
\Text(10,0)[]{$\lambda$}
\Text(170,0)[]{$\bar{\lambda}$}
\Text(90,0)[]{$\Gamma_{\mu\nu}$}
\Text(90,80)[]{$ M$}
\ArrowLine(230,10)(309,10)
\ArrowLine(311,10)(390,10)
\PhotonArc(310,40)(30,0,360){2}{30}
\ArrowArc(310,40)(30,-90,90)
\ArrowArc(310,40)(30,90,270)
\Vertex(310,10){1.5} \Vertex(310,70){2}
\Text(360,40)[]{$\psi^{\mu}_{L}$}
\Text(260,40)[]{$\bar{\psi}^{\nu}_{R}$}
\Text(230,0)[]{$\lambda$}
\Text(390,0)[]{$\bar{\lambda}$}
\Text(310,0)[]{$\Gamma_{\mu\nu}$}
\Text(310,80)[]{$ M$}
\end{picture}
\end{center}
\caption{Gaugino masses induced at 1-loop order, we have introduced $\Gamma_{\mu\nu}=\eta_{\mu\nu}-\frac{1}{3}\gamma_{\mu\nu}$.}
\label{fig:gaugmas4d}
\end{figure}
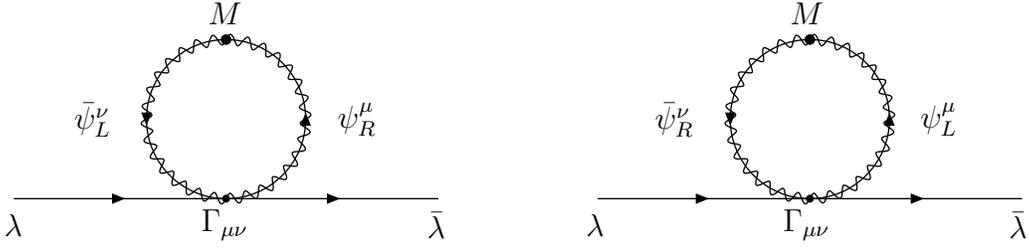

		Let us turn to the five dimensional case. In a most general situation in the presence of arbitrary boundary terms one does not know 
the exact structure of coupling of the five dimensional supergravity to branes\footnote{See \cite{Rattazzi:2003rj} for a discusion of brane-bulk couplings.}. However, one can expect that the effective theory should 
reconstruct the four-dimensional structure described above, with the modification, that the fields that enter (\ref{couplgg}) are the fermionic modes 
which have a nonzero amplitude on the brane with the gauge sector in question. Taking the general solution for the gravitini (\ref{solgravitini3}), one can check that only one half of the fermionic degrees of freedom
 couples to the specific brane. For example, the combination that couples to the brane at the point $y=0$   is given by
		 \begin{equation}
		 \psi_{\mu\,R}^{0\,(n)}=\frac{1}{\sqrt{2}}\left(\tilde{\psi}^{(n)}_{\mu\,R}+\tilde{\chi}^{(n)}_{\mu\,R}\right)\ ,\qquad 
		\psi_{\mu\,L}^{0\,(n)}=\frac{1}{\sqrt{2}}\left(\tilde{\psi}^{(n)}_{\mu\,L}-\tilde{\chi}^{(n)}_{\mu\,L}\right)\ .
		 \end{equation}
		The orthogonal combination
		\begin{equation}
		 \psi_{\mu\,R}^{\pi\,(n)}=\frac{1}{\sqrt{2}}\left(\tilde{\psi}^{(n)}_{\mu\,R}-\tilde{\chi}^{(n)}_{\mu\,R}\right)\ ,\qquad 
		\psi_{\mu\,L}^{\pi\,(n)}=\frac{1}{\sqrt{2}}\left(\tilde{\psi}^{(n)}_{\mu\,L}+\tilde{\chi}^{(n)}_{\mu\,L}\right)\ ,
		 \end{equation}
		 decouples from the brane. Notice that the gravitini in the new basis form {\em Majorana} spinors such that right and 
left handed combinations have  opposite charges under the $U(1)_R$ symmetry preserved by boundary condition given by (\ref{unbgen0}). Of course, 
spinors in the new basis are not eigenstates of the mass matrix and the mass  terms in the Lagrangian take the form
		\begin{eqnarray}
		&&{\cal L}_{mass}
        =-\frac{1}{2}\sum_n \Big(\bar{M}\,\bar{\psi}^{0\, (n)}_\mu\gamma^{\mu\nu}\psi^{0\, (n)}_{\nu}
        +\bar{M}\,\bar{\psi}^{\pi\, (n)}_\mu\gamma^{\mu\nu}\psi^{\pi\, (n)}_{\nu}\nonumber\\
		&&\qquad\qquad\qquad\quad-\bar{m}_n\,\bar{\psi}^{0\, (n)}_\mu\gamma^{\mu\nu}\psi^{\pi\, (n)}_{\nu}
        -\bar{m}_n\,\bar{\psi}^{\pi\, (n)}_\mu\gamma^{\mu\nu}\psi^{0\, (n)}_{\nu}\Big)\ .
		\end{eqnarray}		
The masses
		\begin{equation}
		\bar{m}_n=\frac{{\rm n+\frac{1}{2}}}{r_c}\ ,
		\end{equation}
mix $\psi^{0\,(n)}_\mu$ and $\psi^{\pi\,(n)}_\mu$ states and do not violate the $U(1)_R$ symmetry, since the left/right handed component of $\psi^{0\,(n)}_\mu$ has the same charge as the right/left handed component of $\psi^{\pi\,(n)}_\mu$.  The terms which depends on $\alpha_{0/\pi}$
form  {\em Majorana} mass terms  that have  the same form at each Kaluza-Klein level:  
		\begin{equation}
		\bar{M}=\frac{1}{r_c}\left(\frac{1}{2}-\frac{1}{\pi}\arctan\left(\frac{\alpha_0+\alpha_\pi}{1-\alpha_0\alpha_\pi}
		\right)\right) \ ,
		\end{equation}
		and break   the $U(1)_R$ symmetry. One should note that in the flipped limit $\bar{M}=0$, which agrees with the fact that  the $U(1)_R$ symmetry remains unbroken in that case. 
		
		Let us consider a vector supermultiplet localized on the brane at $y=0$. In the effective four-dimensional theory only one half of the gravitini degrees of freedom couples
 to this supermultiplet, precisely  the same modes which couple to the brane ($\psi^{0\,(n)}_\mu$). To be able to close the diagrams  that produce one-loop effective masses for the gaugini, see Figure \ref{fig:gaugmas4d}, 
 one needs a nonzero $\bar{M}$.

        \section{Super-higgs effect in the presence of flipped boundary conditions} \label{sectionflip}
	In this section we shall present in some detail  the super-higgs mechanism arising in supergravity spontaneously broken
	by  non-trivial boundary conditions (the Scherk-Schwarz mechanism). We shall explicitly show that the longitudinal
	 degrees of freedom for massive gravitini come from the super-higgs mechanism that occurs at
    each level of the Kaluza-Klein tower. The fifth component of the five-dimensional gravitini
     is absorbed by the four-dimensional gravitini. We shall avoid artificial diagonalization of infinitely dimensional matrices known from the earlier work. 
Our final results agree for instance with those of \cite{Bagger:2001ep} when they overlap. To start with, let us concentrate on the gravitini equation of motion
	in the bulk: 
    \begin{eqnarray} \label{equationosmo}
    &&\gamma^{\mu\nu\rho}\partial_\nu\Psi_{\rho}^{A}+\gamma^5\gamma^{\mu\nu}\partial_\nu\Psi_{5}^{A}-
	\gamma^5\gamma^{\mu\nu}\partial_5\Psi_{\nu}^{A}=0\ ,\nonumber\\ 
	&&\gamma^5\gamma^{\mu\nu}\partial_\mu\Psi_{\nu}^{A}=0\ .
    \end{eqnarray}
    We performe the calculation for the flipped supergravity ($\alpha_0=1/\alpha_\pi$), hence the boundary conditions
    take the form
     \begin{eqnarray} \label{warunkibrzeggravsh}
      &&\epsilon^{-1}(y) \delta(y) \gamma_5(\Psi_-)_\mu^A
  =-\delta(y)\alpha_{0}\sigma_1(\Psi_+)_\mu^A\
  ,\nonumber\\
    &&\epsilon^{-1}(y) \delta(y-\pi r_c) \gamma_5(\Psi_-)_\mu^A
  =\delta(y-\pi r_c)(1/\alpha_{0})\sigma_1(\Psi_+)_\mu^A\ .
      \end{eqnarray}  
      One can easily find solutions:
      \begin{eqnarray} \label{solgravitinish}
        &(\Psi_{+})_{\mu}^A&=\sum_n A^{(n)}\left(\cos(m_n|y|)\left(\begin{array}{c}\psi^{(n)}_{\mu\,R}\\
        \chi^{(n)}_{\mu\,L}\end{array}\right)^A+\alpha_0\sin(m_n|y|)
		\left(\begin{array}{c}\chi^{(n)}_{\mu\,R}\\
        \psi^{(n)}_{\mu\,L}\end{array}\right)^A\right)\nonumber\\
        &(\Psi_{-})_{\mu}^A&=\epsilon(y) \sum_n A^{(n)} \left(\sin(m_n|y|)\left(\begin{array}{c}-\psi^{(n)}_{\mu\,L}\\
        \chi^{(n)}_{\mu\,R}\end{array}\right)^A+ \alpha_0 \cos(m_n|y|)
		\left(\begin{array}{c}\chi^{(n)}_{\mu\,L}\\
        -\psi^{(n)}_{\mu\,R}\end{array}\right)^A\right)\nonumber\\
		&(\Psi_{+})_{5}^A&=\epsilon(y)\sum_n A^{(n)}\left(\sin(m_n|y|)\left(\begin{array}{c}\psi^{(n)}_R\\
        \chi^{(n)}_L\end{array}\right)^A-\alpha_0\cos(m_n|y|)
		\left(\begin{array}{c}\chi^{(n)}_R\\
        \psi^{(n)}_L\end{array}\right)^A\right)\nonumber\\
        &(\Psi_{-})_{5}^A&= \sum_n A^{(n)} \left(\cos(m_n|y|)\left(\begin{array}{c}\psi^{(n)}_L\\
        -\chi^{(n)}_R\end{array}\right)^A+ \alpha_0 \sin(m_n|y|)
		\left(\begin{array}{c}\chi^{(n)}_L\\
        -\psi^{(n)}_R\end{array}\right)^A\right)\ ,
        \end{eqnarray}
        where $\psi^{(n)}_\mu$, $\chi^{(n)}_\mu$ and
        $\psi^{(n)}$, $\chi^{(n)}$ denote
        4d gravitini and fermions in the flat space, which satisfy
        \begin{eqnarray} \label{equationofm}
        &&\gamma^{\mu\rho\nu}\partial_{\rho}\psi^{(n)}_{\nu}-\gamma^{\mu\rho}\partial_{\rho}\psi^{(n)}
		- m_n\,\gamma^{\mu\nu}\psi^{(n)}_{\nu}=0\ ,\nonumber\\
		&&\gamma^{\mu\rho\nu}\partial_{\rho}\chi^{(n)}_{\nu}-\gamma^{\mu\rho}\partial_{\rho}\chi^{(n)}
		- m_n\,\gamma^{\mu\nu}\chi^{(n)}_{\nu}=0\ ,
		\end{eqnarray}
		 with the additional conditions $\gamma^{\mu\nu}\partial_\mu\psi^{(n)}_{\nu}=\gamma^{\mu\nu}\partial_\mu\chi^{(n)}_{\nu}=0$. 
		 One can easily find the normalization constant: $A^{(n)}=1/\sqrt{\pi r_c (1+\alpha_0^2)}$.
		 
       The boundary conditions (\ref{warunkibrzeggravsh}) imply
		 the quantization of the masses:
		\begin{equation}
		m_n=\frac{1}{r_c}\left({\rm n}+\frac{1}{2}\right)\ ,\quad {\rm for}\;\;{\rm n}\in {\bf N}\ .
		\end{equation}

		Let us investigate the effective four-dimensional theory. Putting the solutions 
		(\ref{solgravitinish}) into the supergravity  action (\ref{sugravaction}) leads to
		the following four-dimensional Lagrangian describing gravitini
		\begin{eqnarray}
		{\cal L}_{3/2}=&&-\frac{1}{2}\int_{-\pi r_c}^{\pi r_c}\bar{\Psi}^A_\alpha\gamma^{\alpha\beta\gamma}\partial_\beta\Psi_{\gamma
        A}-\frac{1}{2}e_5^{-1}e_4\alpha_0\bar{\Psi}_\mu^A\gamma^{\mu\nu}
		(\sigma_1+{\rm i}\gamma_5\sigma_2)_{A}^{\;B} \Psi_{\nu B}\Big|_{y=0}\nonumber\\
		&&\;\,\qquad\qquad\qquad\qquad\qquad-\frac{1}{2\alpha_0}e_5^{-1}e_4\bar{\Psi}_\mu^A\gamma^{\mu\nu}
		(\sigma_1+{\rm i}\gamma_5\sigma_2)_{A}^{\;B} \Psi_{\nu B}\Big|_{y=\pi r_c}\nonumber\\
        =&&-\frac{1}{2}\int_{-\pi r_c}^{\pi r_c}\left(\bar{\Psi}^A_\mu\gamma^{\mu\nu\rho}\partial_\nu\Psi_{\rho
        A}+\bar{\Psi}^A_5\gamma^5\gamma^{\mu\nu}\partial_\mu\Psi_{\nu
        A}+\bar{\Psi}^A_\mu\gamma^5\gamma^{\mu\nu}\partial_\nu\Psi_{5
        A}-\bar{\Psi}^A_\mu\gamma^5\gamma^{\mu\nu}\partial_5\Psi_{\nu
        A}\right)\nonumber\\
        &&-\frac{1}{2}e_5^{-1}e_4\alpha_0\bar{\Psi}_\mu^A\gamma^{\mu\nu}
		(\sigma_1+{\rm i}\gamma_5\sigma_2)_{A}^{\;B} \Psi_{\nu B}\Big|_{y=0}
		-\frac{1}{2\alpha_0}e_5^{-1}e_4\bar{\Psi}_\mu^A\gamma^{\mu\nu}
		(\sigma_1+{\rm i}\gamma_5\sigma_2)_{A}^{\;B} \Psi_{\nu B}\Big|_{y=\pi r_c}\nonumber\\
        =&&-\frac{1}{2}\sum_n \left(\bar{\psi}^{(n)}_\mu\gamma^{\mu\nu\rho}\partial_\nu\psi^{(n)}_{\rho}
        +\bar{\chi}^{(n)}_\mu\gamma^{\mu\nu\rho}\partial_\nu\chi^{(n)}_{\rho}
		-m_n\,\bar{\psi}^{(n)}_\mu\gamma^{\mu\nu}\psi^{(n)}_{\nu}
        -m_n\,\bar{\chi}^{(n)}_\mu\gamma^{\mu\nu}\chi^{(n)}_{\nu}\right)\nonumber\\
        &&-\frac{1}{2}\sum_n \left(\bar{\psi}^{(n)}\gamma^{\mu\nu}\partial_\mu\psi^{(n)}_{\nu}
        +\bar{\chi}^{(n)}\gamma^{\mu\nu}\partial_\mu\chi^{(n)}_{\nu}
		-\bar{\psi}^{(n)}_\mu\gamma^{\mu\nu}\partial_\nu\psi^{(n)}
        -\bar{\chi}^{(n)}_\mu\gamma^{\mu\nu}\partial_\nu\chi^{(n)}\right)\ .
		\end{eqnarray}
		The variational principle leads to the four-dimensional equation of motion (\ref{equationofm}).
		  One can remove from the Lagrangian $\psi^{(n)}$ and $\chi^{(n)}$ fields by the following redefinition:
		  \begin{eqnarray}\label{transformsh}
		  &&\psi^{(n)}_\mu\longrightarrow \psi^{(n)}_\mu-\frac{1}{m_n}\,\partial_\mu \psi^{(n)}\nonumber\\
			&&\chi^{(n)}_\mu\longrightarrow \chi^{(n)}_\mu-\frac{1}{m_n}\,\partial_\mu \chi^{(n)}\ .
		  \end{eqnarray}
		Also the equation (\ref{equationofm}) reduces to the standard Rarita-Schwinger equation
		 \begin{eqnarray} 
        &&\gamma^{\mu\rho\nu}\partial_{\rho}\psi^{(n)}_{\nu}
		- m_n\,\gamma^{\mu\nu}\psi^{(n)}_{\nu}=0\nonumber\\
		&&\gamma^{\mu\rho\nu}\partial_{\rho}\chi^{(n)}_{\nu}
		- m_n\,\gamma^{\mu\nu}\chi^{(n)}_{\nu}=0\ .
		\end{eqnarray}
		
		One should note that the transformations (\ref{transformsh}) are the  part of the supersymmetry
		transformations of the gravitini with the parameters $\psi^{(n)}$ and $\chi^{(n)}$. In the more general case, when one 
		considers the reduction of the full five-dimensional action including the interaction term between gravitini and graviphoton,
		the redefinitions which remove the fermions $\psi^{(n)}$ and $\chi^{(n)}$ from the four-dimensional Lagrangian should also include
		terms with graviphoton in the same manner as they appear in the full supersymmetry transformation of the 
		four-dimensional gravitino.

 \section{Summary}       
The scenarios of split supersymmetry have demonstrated that the current phwenomenological constraints 
can safely be satisfied in models with a large hierarchy between supersymmetry breaking terms. Using  simple
locally supersymmetric five-dimensional models we have demonstrated at field theoretical level how the scenario proposed by Antoniadis and 
Dimopoulos in \cite{Antoniadis:2004dt} realizes such a hierarchy, in fact - an arbitrary hierarchy, between gravitini mass terms and masses of gaugini. 
Indeed, it turns out that for the special choice of boundary conditions realized by a set of brane sources there appears an unbroken 
R-symmetry (with supersymmetry broken at the same time) which forbids gaugino masses while gravitini masses are non-vanishing. 
Departure from this symmetric set of boundary conditions breaks R-symmetry, and gaugino masses can be generated at one-loop order, 
however the magnitude of the resulting soft masses is proportional to the R-symmetry breaking Majorana-type gravitini mass,  
which is continously deformable to zero (at the R-symmetric point). 
In contrast to N=1 supergravity all gravitini are massive but R-symmetry can stay unbroken, since in the limiting case
with boundary sources of supersymmetry breaking absent, the superhiggs effect is contained within the gravitational sector. 
While construction of the working field theoretical extension of the Standard Model 
along the lines discussed here may be a formidable task, the scenario is certainly interesting, as it allows one to avoid constraints imposed by the tight 
framework of N=1 4d supergravity.   

\vskip 1.7cm

\centerline{\Large \bf Acknowledgements}

\vspace*{0.5cm}
\noindent This work was partially supported by the EC 6th Framework
Programme MRTN-CT-2004-503369,  by the Polish State Committee for Scientific
Research grant KBN 1 P03D 014 26 and by POLONIUM 2005.
R.M. gratefully acknowledges financial support from the European Network
for Theoretical Astroparticle Physics (ENTApP), member of ILIAS, EC
contract number RII-CT-2004-506222.


\newpage
\appendix
\noindent{\large \bf Appendix A:  Supergravity on ${\bf S^1}/{\bf Z_2}$}
\setcounter{equation}{0}
\renewcommand{\theequation}{A.\arabic{equation}}
\vspace{0.3cm}

Let us define  five-dimensional, N=2 supergravity  on ${\cal M}_4\times {\bf S^1}/{\bf Z_2}$,
where ${\cal M}_4$ denotes four-dimensional Minkowski space-time. Simple supergravity
multiplet contains: metric tensor (represented by the vielbein $e^m_\alpha$), two gravitini $\Psi^A_\alpha$
and  vector field $A_\alpha$
-- the graviphoton.
The pair of gravitini satisfies symplectic Majorana condition
 $\bar{\Psi}^A\equiv\Psi_A^\dagger\gamma_0=(\epsilon^{AB}\Psi_B)^TC$.
 Five-dimensional  Lagrangian reads
        \begin{eqnarray} \label{sugravaction}
        &{\cal L}_{grav}=&\frac{1}{2}R-\frac{3}{4}{\cal F}_{\alpha\beta}{\cal F}^{\alpha\beta}
        -\frac{1}{2\sqrt{2}}A_\alpha{\cal F}_{\beta\gamma}{\cal F}_{\delta\epsilon}\epsilon^{\alpha\beta\gamma\delta\epsilon}\nonumber
        \\&&-\frac{1}{2}\bar{\Psi}^A_\alpha\gamma^{\alpha\beta\gamma}\partial_\beta\Psi_{\gamma
        A}\nonumber\\
        &&+\frac{3{\rm i}}{8\sqrt{2}}\left(\bar{\Psi}^A_\gamma\gamma^{\alpha\beta\gamma\delta}\Psi_{\delta A}
        +2\bar{\Psi}^{\alpha A}\Psi_{A}^{\beta}\right){\cal F}_{\alpha\beta}\ ,
        \end{eqnarray}
       with supersymmetry transformations
        \begin{eqnarray}
        &&\delta e^m_\alpha=\frac{1}{2}\bar{\eta}^A\gamma^m\Psi_{\alpha A},
        \;\;\delta A_\alpha=-\frac{{\rm i}}{2\sqrt{2}}\bar{\Psi}_{\alpha}^A\eta_A,\nonumber
        \\&&\delta\Psi_{\alpha}^A=\partial_\alpha\eta^A- \frac{{\rm i}}{4\sqrt{2}}
        \left(\gamma_\alpha^{\;\beta\gamma}-4\delta_\alpha^{\;\beta}\gamma^\gamma\right)
        {\cal F}_{\beta\gamma}\eta^A\ .
        \end{eqnarray}
	One should note at this point that the above Lagrangian is invariant under the $SU(2)_R$ symmetry, 
that acts on the symplectic indices. The graviton and the graviphoton form singlets with respect to this symmetry, while 
the pair of gravititni and the parameters of the supersymmetry transformations $\eta^A$ transform as  doublets.

	We pass on to the orbifold ${\bf S^1}/{\bf Z_2}$ by
     identifying $(x_\mu,y)$ with $(x_\mu,-y)$ and choosing the action of ${\bf
     Z_2}$ on the fields. In the bosonic sector we have chosen even parity for
     $e^a_\mu$, $e^5_5$, $A_5$ and odd parity for $e^5_\mu$, $e^a_5$,
     $A_\mu$. In the fermionic sector ${\bf Z_2}$ operators $Q_0$ and $Q_\pi$ acts on the fields as
     follows
     \begin{eqnarray}\label{gbcond}
        &&\Psi^A_\mu(-y)=\gamma_5(Q_0)^A_{\;B}\Psi^B_\mu(y)\ ,\quad\;\;\:\Psi^A_\mu(\pi r_c-y)=\gamma_5(Q_\pi)^A_{\;B}\Psi^B_\mu(\pi r_c+y)\
        ,\nonumber\\
        &&\Psi^A_5(-y)=-\gamma_5(Q_0)^A_{\;B}\Psi^B_5(y)\ ,\quad\Psi^A_5(\pi r_c-y)=-\gamma_5(Q_\pi)^A_{\;B}\Psi^B_5(\pi r_c+y)\
        ,\\
        &&\eta^A(-y)=\gamma_5(Q_0)^A_{\;B}\eta^B(y)\ ,\quad\;\;\;\;\:\eta^A(\pi r_c-y)=\gamma_5(Q_\pi)^A_{\;B}\eta^B(\pi r_c+y)\
        .\nonumber
     \end{eqnarray}
     The symplectic Majorana condition and the normalization $(Q_{0,\pi})^2=1$
     imply that ${\bf Z_2}$ operators can be written as the following linear combinations of the Pauli
     matrices:
     $Q_{0,\pi}=(q_{0,\pi})_i \sigma^i$, where $(q_{0,\pi})_i$ form
     real unit vector. In general, one can choose different $Q_i$ operators at each orbifold fixed point ($y=0$ or
	$y=\pi r_c$). 

\newpage

\end{document}